\documentclass[pra,twocolumn]{revtex4}
\usepackage{graphicx}

\begin{document}
\title{Thermal entanglement in superconducting quantum-interference-device qubits coupled to cavity field\footnote{Supported by the National Fundamental
Research Program Grant No. 2001CB309310, the National Natural
Science Foundation under Grant Nos. 10325523, 90203018 and
10075018, the key projects of the Education Ministry of China, and
the Educational Committee of Human Province under Grant Nos. 03094
and 02A026.}}
\author{PAN Hai-Zhong  and KUANG Le-Man\footnote{Corresponding
author. Email: lmkuang@hunnu.edu.cn}}
\address{Department of Physics, Hunan Normal University, Changsha 410081, China\ }

\begin{abstract}
In this paper, we investigate thermal entanglement in a
superconducting-quantum-interference-device qubit coupled to a
cavity field. We show that the entanglement can be manipulated by
varying temperature and an effective controlling parameter $B$
which depends on the external field  and characteristic parameters
of the system. We find that there exists a critical value of the
controlling parameter $B_c$. Under a fixed temperature, increasing
$B$ can increase entanglement in the regime of $ B<B_c$, while the
entanglement decreases with increasing $B$ in the regime of $
B>B_c$.

\vspace{0.5cm}
 \noindent PACS number(s): 03.67.-a, 03.65.Ud, 85.25.Dq

\end{abstract}

\maketitle


For decades, quantum entanglement has been the focus of much work
in the foundations of quantum mechanics. Beyond this fundamental
aspect, creating and manipulating of entangled states are
essential for quantum information applications. Hence, quantum
entanglement \cite{cai} has been viewed as an essential resource
for quantum information processing. In recent years, much
attention has been paid to the presence of entanglement in
condensed-matter systems at finite temperatures. The state of a
typical condensed-matter system at thermal equilibrium
(temperature $T$) is described by the density operator
$\rho=e^{-\beta H}/Z$ where $H$ is the Hamiltonian, $Z=tre^{-\beta
H}$ is the partition function, and $\beta=1/kT$ where $k$ is
Boltzmann¡¯s constant. The entanglement associated with the
thermal state $\rho$ is referred to as thermal entanglement
\cite{arn}.

Superconducting Josephson devices (SJD's) are  simple but
realistic and extensively studied solid systems. In fact, they may
be thought of as two-state systems realizing the elementary unit
of quantum information, known as a qubit. Moreover, Josephson
devices \cite{mak,pas,ber,yu,vio,shn,nak,mig,liu,kua,kuan} can be
scaled up to a large number of qubits and their dynamics may be
controlled by externally applied voltages and magnetic fluxes.
Superconducting devices such as Cooper pair boxes, Josephson
junctions, or superconducting quantum interference devices
(SQUID's) have been thus proposed and used as basic elements for
the practical realization of quantum gates.

It is an interesting topic to study thermal entanglement in
systems involving SJD's, such as SQUID's. As we know, so far no
study on thermal entanglement of Josephson qubit systems has been
reported. The purpose of this paper is to investigate thermal
entanglement of such a Josephson qubit system which consists of a
flux qubit (an rf-SQUID) coupled to a single-mode quantized
electromagnetic field of a resonant cavity.

The system which we consider is an rf-SQUID coupled to a
single-mode electromagnetic field of a high-$Q$ resonant cavity.
The rf-SQUID is a superconducting loop interrupted by a Josephson
junction with the junction capacitance $C$ and Josephson coupling
energy $E_J$. The Hamiltonian of the rf-SQUID \cite{mak} is given
by
\begin{eqnarray}
\label{e1}
H_S=\frac{Q^{2}}{2C}-E_{J}\cos(2\pi\frac{\phi}{\phi_{0}})+\frac{(\phi-\phi_{x})^{2}}{2L},
\end{eqnarray}
where $\phi_x$ is an externally applied dc flux, $\phi_0=h/2e$ the
flux quantum, $L$ the self-inductance of the loop, the charge $Q$
on the junction capacitance and the flux $\phi$ in the loop are
canonically conjugate operators obeying the commutation relation
$[\phi, Q]=i\hbar$.

If the self-inductance is large, such that
$E_J>\phi_0^2/(4\pi^2L)$ and the externally applied flux $\phi_x$
is close to $\phi_0/2$, the last two terms in (\ref{e1}) form a
double-well potential near $\phi=\phi_0/2$. Actually, the form of
the potential in Hamiltonian (\ref{e1}) can be tuned by changing
the external dc magnetic flux $\phi_x$ applied to the loop to the
case where the two lowest-energy wells are degenerate. This case
occurs for $\phi_x=\phi_0/2$. These two degenerate minima
correspond to the clockwise and counterclockwise sense of rotation
of the supercurrent in the loop. The two relative ground states
are localized flux states, hereafter denoted $|L\rangle$ and
$|R\rangle$. At low temperatures only the lowest states in the two
wells contribute. Hence the rf-SQUID reduces to an effective
two-state system \cite{mig} with the Hamiltonian
\begin{equation}
\label{e2}
H_S=-\frac{\hbar}{2}\Delta\sigma_{x}+\frac{\hbar}{2}\epsilon\sigma_{z},
\end{equation}
where the diagonal term
$\hbar\epsilon=2I_{c}\sqrt{6(\beta-1)}(\phi_{x}-\phi_{0}/2)$ is
the bias, i.e.,  the asymmetry of the double well, $\hbar\epsilon$
can be tuned by the applied flux $\phi_x$. The off-diagonal term
 $\Delta$ describes  the inter-well tunnelling amplitude that depends
on the Josephson coupling energy $E_J$ which may be manipulated by
 a dc flux $\phi_c$ in a dc-SQUID replacing the single
Josephson junction in the rf-SQUID .

\begin{figure} [htbp]
\includegraphics[width=8cm,height=5cm]{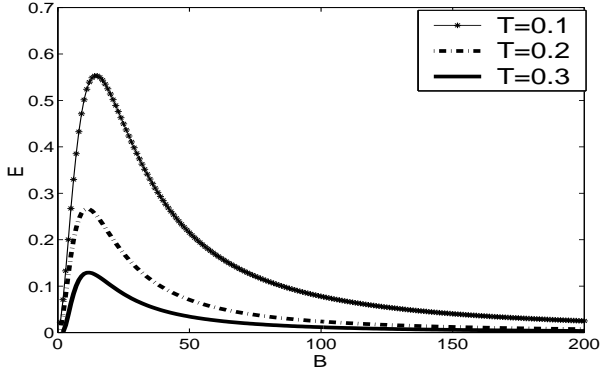}
\caption{Entanglement as a function of the effective controlling
parameter for different values of temperature: $T=0.1$ (star
line), $T=0.2$ (dashed line), and $T=0.3$ (solid line).  Here $B$
is in unit of $10^{-35}$ J s.}
\label{cong:fig1}
\end{figure}

The eigenvectors of Hamiltonian (\ref{e2}) are given by
\begin{eqnarray}
\label{3}
|-\rangle&=&N_{-}[|R\rangle+(\epsilon+\sqrt{\epsilon^{2}+\Delta^{2}})/\Delta|L\rangle],
\\
\label{4}
|+\rangle&=&N_{+}[|R\rangle+(\epsilon-\sqrt{\epsilon^{2}+\Delta^{2}})/\Delta|L\rangle],
\end{eqnarray}
where the normalization constants
$N_{\pm}=(1+[(\epsilon\pm\sqrt{\epsilon^{2}+\Delta^{2}})/\Delta])^{-1/2}$,
and the corresponding eigenvalues are
$E_{\pm}=(\pm\hbar/2)\sqrt{\epsilon^{2}+\Delta^{2}}$.

The Hamiltonian of the single-mode cavity field can be written as
$H_c=\hbar\omega_c(\hat{a}^{\dagger}\hat{a}+1/2)$ where $\omega_c$
is the frequency of the cavity field. One may model the cavity
field as an $LC$ resonator with the resonant frequency
$\omega_c=1/\sqrt{L_cC_c}$ where the capacitive parameter $C_c$
depends on the cavity field frequency via the quantization volume
and on the rf-SQUID geometry \cite{mig}.

The flux qubit and the $LC$-resonator modelling the single-mode
cavity field can be thought of as coupled together inductively.
The interaction Hamiltonian \cite{mig} can be described by
\begin{equation}
\label{e6}
H_I=B(\hat{a}+\hat{a}^{\dagger})(-\epsilon\sigma_{z}+\Delta\sigma_{x}),
\end{equation}
where the controlling parameter $B$ depends on the coupling
strength and the system characteristic parameters, it can be
explicitly expressed as
\begin{equation}
\label{e7}
B=\frac{k}{L}\sqrt{\frac{\hbar}{2\omega_{c}C_{c}}}\frac{\phi_{0}}{\sqrt{\epsilon^{2}+\Delta^{2}}},
\end{equation}
where $k$ is the flux linkage factor with the order of 0.01.

Hence, the total Hamiltonian can be written as
\begin{equation}
\label{e8} H=H_s + H_c + H_I.
\end{equation}

If we focus our attention on  the initial condition $|n
\sigma\rangle=|n\rangle |\sigma\rangle$ such that no more than one
matter-radiation excitation $n+(n_{\sigma}+1/2)$ with
$n_{\sigma}=\pm 1$ is present. Under this initial condition, it
can be shown that most of the evolution occurs within the
low-lying energy subspace of the free Hamiltonian $H_0=H_s+H_c$
spanned by the four states $|n \sigma\rangle$, with $n=0,1$ and
$\sigma=-,+$. In such a four-state subspace, the dynamics system
is governed by the reduced Hamiltonian
\begin{eqnarray}
\label{e9}
H=\left(\begin{array}{lccr}0&-B\epsilon&0&B\Delta\\
-B\epsilon&\hbar\omega_{c}&B\Delta&0\\
0&B\Delta&\hbar\omega_{c}&B\epsilon\\
B\Delta&0&B\epsilon&2\hbar\omega_{c}\end{array}\right).
\end{eqnarray}

\begin{figure} [htbp]
  \includegraphics[width=8cm,height=5cm]{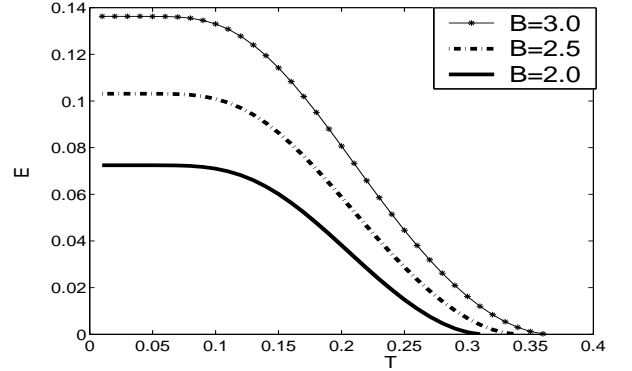}
  \caption{Entanglement as a function of temperature for different values of the controlling parameter: $B=2.0$ (solid line),
$B=2.5$ (dashed line), and $B=3.0$ (star line). Here $B$ is in
unit of $10^{-35}$ J s.}
  \label{cong:fig2}
\end{figure}

We now investigate the thermal entanglement in the combined system
of the rf-SQUID and the cavity field. For the simplicity, we
consider the symmetric case of $\epsilon=0$. In this case,  the
eigenvectors  of Hamiltonian (\ref{e9}) are given by
\begin{eqnarray}
\label{e10} |\psi_{1}\rangle&=&\frac{1}{\sqrt{2}}(|0 +\rangle - |1
-\rangle ), |\psi_{2}\rangle=\frac{1}{\sqrt{2}}(|0 +\rangle + |1
-\rangle ),\\
 |\psi_{3}\rangle&=&N_3(|1 +\rangle - b_+|0 -\rangle),
|\psi_{4}\rangle=N_4(|1 +\rangle - b_-|1 +\rangle), \nonumber
\end{eqnarray}
where the normalization constants $N_{3}=(1+ b_+^2)^{-1/2}$, and
$N_{4}=(1+ b_-^2)^{-1/2}$ with
$b_{\pm}=(\hbar\pm\sqrt{B^{2}+\hbar^{2}})/B$, and corresponding
eigenvalues are given by
\begin{eqnarray}
\label{e11} E_1&=&(\hbar-B)\omega_{c},\hspace{0.3cm}
E_2=(\hbar+B)\omega_{c}\\
E_3&=&(\hbar-\sqrt{B^{2}+\hbar^{2}})\omega_{c},
E_4=(\hbar+\sqrt{B^{2}+\hbar^{2}})\omega_{c}.\nonumber
\end{eqnarray}

We use the entanglement of formation as the entanglement measure
for a state described by a density operator $\rho$. To calculate
this measure, starting from the density matrix , one can introduce
the product matrix $R$ associated with the density matrix and its
time-reversed matrix $
R=\rho(\sigma_{1}^{y}\bigotimes\sigma_{2}^{y})\rho^{*}(\sigma_{1}^{y}\bigotimes\sigma_{2}^{y})$
\cite{woo}. Then concurrence is defined as $
C=max\{\lambda_{1}-\lambda_{2}-\lambda_{3}-\lambda_{4},0\}$ where
the $\lambda_{i}$  are the square roots of the eigenvalues of $R$,
in decreasing order. In this method the standard basis, $|0
+\rangle, |0 -\rangle, |1 +\rangle, |1 -\rangle$ must be used. The
entanglement of formation is a strictly increasing function of
concurrence; thus there is a one-to-one correspondence. The amount
of entanglement is given by entropy defined as
\begin{equation}
\label{e14}
 E=h\left(\frac{1}{2}\left[1+\sqrt{1-C^{2}}\right]\right)
\end{equation}
where $h(x)=-xlog_{2}x-(1-x)log_{2}(1-x)$.

For the composite system  of  the rf-SQUID and the cavity field in
thermal equilibrium  at temperature $T$, making use of the
eigenvectors and eigenvalues given in expressions (\ref{e10}) and
(\ref{e11}), we can get the corresponding density operator $
\rho(T)=\sum_{i=1}^{4}(e_i|\psi_{i}\rangle\langle\psi_{i}|)/$
where $e_i=\exp{(-\beta E_{i})} (i=1,2,3,4)$ and
$Z=\sum_{i=1}^{4}e_i$ is the partition function of the composite
system with $\beta=1/kT$ with $k$ being the Boltzmann's constant.
In the standard basis, $\{|+ 0\rangle, |+ 1\rangle, |- 0\rangle,
|- 1\rangle\}$, the density operator can be written as
\begin{eqnarray}
\label{e17}
\rho(T)=\frac{1}{Z}\left(\begin{array}{lccr}u&0&0&v\\0&x&w&0\\
0&w&y&0\\v&0&0&u\end{array}\right),
\end{eqnarray}
where the nonzero matrix elements are given by
\begin{eqnarray}
\label{e18} u&=&\frac{1}{2}\left(e_2 + e_1\right), \hspace{0.2cm}
v=\frac{1}{2}\left(e_2 - e_1\right), \nonumber\\
x&=&\frac{b_+^2e_3}{1+b_+^2} + \frac{b_-^2e_4}{1+b_-^2},
\hspace{0.2cm}  y=\frac{e_3}{1+b_+^2} +
\frac{e_4}{1+b_-^2}, \nonumber\\
w&=&-\frac{b_+e_3}{1+b_+^2} - \frac{b_-e_4}{1+b_-^2},
\end{eqnarray}
where we have set $b_{\pm}=\left(\hbar \pm
\sqrt{B^{2}+\hbar^{2}}\right)/B$.

Then the product matrix $R$ becomes
\begin{eqnarray}
\label{e20}
R=\frac{1}{Z^2}\left(\begin{array}{lccr}u^2+v^2&0&0&2uv\\0&w^2+xy&2wx&0\\
0&2wy&w^2+xy&0\\2uv&0&0&u^2+v^2\end{array}\right),
\end{eqnarray}
which has the following eigenvalues
\begin{eqnarray}
\label{e21}
r_1&=&(w+\sqrt{xy})^2/Z^2, \hspace{.5cm} r_2=(w-\sqrt{xy})^2/Z^2, \nonumber\\
r_3&=&(u+v)^2/Z^2, \hspace{.5cm} r_4=(u-v)^2/Z^2.
\end{eqnarray}


In order to calculate the amount of entanglement, we have to know
which one among the square roots of the eigenvalues of $R$ is the
largest. This depends on the values of the controlling parameter
and temperature. Hence, it is difficult to analytically  to
calculate the amount of entanglement, and we turn to numerical
investigation. In Figure 1 we have plotted how entanglement
between the rf-SQUID qubit and the cavity field varies with the
controlling parameter $B$ for different values of temperature $T$.
From Fig. 1 we can see that (i) for every given temperature,
entanglement exhibits a peak, which corresponds to the critical
value of the controlling parameter. Entanglement increases with
increasing $B$ in the controlling regime of $B<B_c$, while
entanglement decreases with  increasing $B$ in the regime of
$B>B_c$; (ii) entanglement at the critical point of the
controlling parameter $B_c$ increases while the  critical point
may move to the left  with increasing temperature; (iii) in the
limit of $B\rightarrow\infty$, entanglement approaches zero which
means that the composite system disentangled. This large $B$
behavior can be obtained analytically. Indeed, when $B\gg\hbar$,
we have $b_{\pm}=\pm 1$,  $E_1\approx E_3 $ and $E_2\approx E_4$.
In this case, $\sqrt{r_2}$ is the largest one among the square
roots of the eigenvalues of the product matrix $R$, and the
concurrence $C=max\left\{-e_2/(e_1+e_2),0\right\}=0$ which implies
that entanglement vanishes $E=0$.

Finally we look to the dependence of entanglement on temperature.
In Figure 2 we have plotted the variation of entanglement with
temperature for different fixed values of the controlling
parameter $B$ in the regime below the critical value $B_c$. Figure
2 indicates that for every fixed value of $B$ entanglement
decreases slowly  with increasing temperature in the low
temperature regime of $T<0.15$ while entanglement drops rapidly in
the middle temperature regime of $0.15<T<0.35$, and
disentanglement appears in the high temperature regime of
$T>0.35$. From Fig. 2 we can also see that entanglement increases
with enhancing the value of the controlling parameter $B$.

In summary, we have investigated thermal entanglement in a SQUID
qubit coupled to a cavity field. We have shown that the
entanglement can be manipulated by varying temperature and an
effective controlling parameter $B$ which depends on the external
field  and characteristic parameters of the system. In general,
thermal entanglement increases with increasing the controlling
parameter and decreases with increasing temperature. We have found
that there exists a critical value of the controlling parameter
$B_c$. Thermal entanglement between the SQUID and the cavity field
exhibits different properties below and above the critical value.
Under a fixed temperature, increasing $B$ can increase
entanglement below the critical value, while the entanglement
decreases with increasing $B$ above the critical value. We hope
that the finding of the existence of the critical value of the
controlling parameter is useful to realize quantum information
processing by using SQUID qubits.

\end{document}